\documentclass[12pt]{article}
\usepackage{cite}
\usepackage{amstex}
\usepackage{amssymb}
\newenvironment{myitemize}{%
      \begin{itemize}}{\end{itemize}}
\renewcommand{\theequation}{\thesection.\arabic{equation}}

\begin{document}
\begin{titlepage}
\title{Not all adiabatic vacua are physical states}

\makeatletter
\author{J. Lindig\thanks{e-mail: lindig@itp.uni-leipzig.de}\\
{\it \small University of Leipzig, Institute for Theoretical Physics},\\ 
{\it \small Augustusplatz 10, 04109 Leipzig, Germany}}

\makeatother
\maketitle


\begin{abstract}
Adiabatic vacua are known to be Hadamard states. We show, however that 
the energy-momentum tensor of a linear Klein-Gordon field on 
Robertson-Walker spaces developes a generic singularity 
on the initial hypersurface if the adiabatic vacuum is of order less 
than four. Therefore, adiabatic vacua are physically reasonable only if 
their order is at least four. 

A certain non-local large momentum expansion of the mode functions has 
recently been suggested to yield the subtraction terms needed to remove 
the ultraviolet divergences in the energy-momentum tensor. We find 
that this scheme fails to reproduce the trace anomaly and therefore is 
not equivalent to adiabatic regularisation.
\end{abstract}

\vfil
\thispagestyle{empty}
\end{titlepage}

\pagestyle{myheadings}
\markright{Not all Adiabatic Vacua $\cdots$}
\allowdisplaybreaks

\section{Introduction}\label{sec1}
\setcounter{equation}{0}

The semiclassical theory of quantised fields propagating on a curved
(globally hyperbolic) spacetime does not provide a principle of how
to choose a vacuum state. In the absence of isometries the vacuum
state cannot be associated with such symmetries of the underlying spacetime. 
Instead, physically reasonable states (of linear fields) are required to be 
Hadamard states, i.e. the corresponding two-point functions have to possess 
the Hadamard singularity structure in order to allow for standard
renormalisation \cite{fulling,kaywald}.

The proper choice of an initial state is not only essential for a 
consistent formulation of quantum field theory on curved spacetimes.
In the context of concrete applications the dependence of the physical 
effects on the initial state becomes an equally significant aspect.
This question arises, for example, in inflationary cosmology where
particle creation and back reaction due to quantum fields play an 
important role. The interest in the consideration of these effects has
recently been intensified in connection with the theory of reheating 
after inflation \cite{kofstarlind} (a discussion of Hadamard 
states in this case is appropriate because the quantum fluctuations 
satisfy linear equations of motion in the mean field approximation 
\cite{motthab94}).

The concept of adiabatic vacua was introduced by Parker in order to 
account for particle creation in an expanding universe \cite{parker}.
The physical motivation behind the adiabatic particle picture is that 
it most closely resembles the particle concept of a static universe
during an expansion. The notion of adiabatic vacuum states was put
on a solid mathematical basis by Roberts and L\"uders \cite{RobLue}
who also suggested that adiabatic vacua and Hadamard states define
the same class of physical states on the cosmologically relevant 
Robertson-Walker spaces. Indeed, both concepts are intimately related.
Najmi and Ottewill \cite{najmotte} derived the leading asymptotic momentum 
behaviour of a second order adiabatic vacuum as a {\it necessary} condition for
Hadamard states on a quasi-euclidean space ($\kappa =0$). Using Fourier 
analysis they compared the symmetric two-point function and its first
derivative with the Hadamard series on the initial hypersurface.
A related analysis can be found in \cite{mazzi}. Recently, Junker has 
succeeded in showing that in fact all adiabatic vacua are 
Hadamard states \cite{junker}. His proof exploits methods of the theory
of pseudodifferential operators and wavefront sets on manifolds.

The expectation value of the energy-momentum tensor rather than the 
two-point function is the essential physical quantity to be considered
because it determines the back reaction effect on the gravitational
field via the semi-classical Einstein equations
\begin{equation}\label{e11}
G_{\mu \nu}=-8\pi G \langle T_{\mu \nu} \rangle.
\end{equation}
The energy-momentum tensor involves second derivatives of the two-point
function. However, the method of \cite{najmotte} could not be generalized
to the case of a second  derivative. So when considering the energy-momentum 
tensor one might expect to find further constraints on the physically 
admissible states. 

It has recently been shown \cite{baacke1} that the expectation value of the
energy-momentum tensor in a conformal-like initial state 
(see eq. (\ref{e35}) below) developes an initial singularity, i.e. the limit
$\eta\to \eta_0$ does not exist ($\eta$ is the conformal time parameter).
Since an initially singular energy-momentum-tensor does not satisfy Wald's 
axioms \cite{fulling} such states should not be considered physically 
reasonable. 

In the present paper we are concerned with the question whether adiabatic
states of linear Klein-Gordon fields on Robertson-Walker spaces (with 
arbitrary spatial curvature) can lead to initial singularities as well. 
We show that the order of an adiabatic vacuum must not be less than four for 
the energy-momentum tensor to be finite on the initial hypersurface. 
As a primary new result we find that even though {\it all} adiabatic vacua are 
Hadamard states \cite{junker} they are physically admissable only if their 
order is four at least. 

In line with our result, the adiabatic particle picture developed in 
\cite{carmen} shows that for adiabatic vacua of order four or higher the 
energy-momentum tensor splits naturally in a local part containing all the
ultraviolet divergences and a finite, non-local piece that can be viewed as
being due to particle production. 

In the derivation of the condition on the adiabatic order we employ a 
non-local large momentum expansion of the conformal-like  mode functions 
(see the appendix) that has similarily been used in \cite{baacke1,davunr,ringw}. 
We show that the subtraction of the leading terms of this expansion as suggested in 
\cite{baacke2} is not equivalent to adiabatic regularisation on 
Robertson-Walker spaces because it fails to reproduce the trace anomaly.
Besides, our proof reveals that the construction of states suggested in 
\cite{baacke1} effectively determines a fourth order adiabatic vacuum.

The paper is organized as follows. In section 2 we review the basic
elements of scalar field quantisation in Robertson-Walker spaces
including adiabatic regularisation as far as necessary and give the
definition of adiabatic states following \cite{RobLue}. In section 3
we show that the adiabatic order of the state must not be less than
four to result in an initially well behaved energy-momentum tensor.
We conclude the paper with a brief summary and a technical appendix. 
Our metric convention is $g_{\mu\nu}={\rm diag}(1,-1,-1,-1)$ and we use 
units such that 
$\hbar =c=1$.

\section{Quantum fields on Robertson-Walker spaces}\label{sec2}
\setcounter{equation}{0}

The Robertson-Walker metric is given by
\begin{equation}\label{e21}
ds^2=a^2(\eta)\left[d\eta^2-h_{ik}dx^idx^k\right],
\end{equation}
where $h_{ik}$ denotes the metric of a $3$-space of constant curvature
$\kappa = -1, 0, +1$ for an open, flat and closed universe, respectively.

The free scalar field satisfies the Klein-Gordon equation
\begin{equation}\label{e22}
\left(\Box +m^2+\xi R\right)\varphi(x)=0.
\end{equation}
The symmetry of the Robertson-Walker metric allows for separating 
variables in eq. (\ref{e22}) and the scalar field can be decomposed as
\begin{equation}\label{e23}
\varphi(x)=\frac{1}{a(\eta)}\int d\tilde{\mu}({\bf k})
\left[f_k(\eta)\Phi_k({\bf x})a_k
+f^*_k(\eta)\Phi^*_k({\bf x})a_k^{\dagger}\right],
\end{equation}
where the creation and annihilation operators $a_k^{\dagger}, a_k$ obey 
the usual commutation relations. The $\Phi_k({\bf x})$ are the 
eigenfunctions of the Laplace-Beltrami operator on the $3$-space of 
constant curvature 
\begin{equation}\label{e24}
\Delta^{(3)}\Phi_k({\bf x})=-\left(k^2-\kappa\right)\Phi_k({\bf x})
\end{equation}
and $d\tilde{\mu}({\bf k})$ is the
measure on the corresponding set of quantum numbers (for details see 
\cite{bunch}). The time-dependent part of the mode function satisfies
the oscillatory equation
\begin{equation}\label{e25}
f_k^{\prime\prime}(\eta)+\Omega_k^2(\eta)f_k(\eta)=0.
\end{equation}
The frequency $\Omega_k(\eta)$ is given by
\begin{equation}\label{e26}
\Omega_k^2(\eta)=k^2+a^2\left(m^2-\Delta\xi R\right)\stackrel{\rm def}{=}
 			 \omega_k^2-q(\eta)\stackrel{\rm def}{=}               
			  k^2+M^2(\eta)
\end{equation}
with $\omega_k^2=k^2+m^2a^2$ and $\Delta\xi =1/6-\xi$. A complete set of mode 
solutions to eq. (\ref{e25}) is specified by imposing initial conditions 
$f_k(\eta_0)$, $f'_k(\eta_0)$ on a Cauchy surface $\eta =\eta_0$. This 
corresponds to the choice of a homogeneous vacuum state. 

We now give the definition of adiabatic vacua following \cite{RobLue}. 
Substituting the WKB ansatz
\begin{eqnarray}
\label{e27}
\tilde f_k(\eta) =  {1 \over \sqrt{2 W_k(\eta)}} 
{\rm exp}\left[-i{\int_{\eta_0}^\eta d\eta' W_k(\eta')}\right] 
\end{eqnarray}
into (\ref{e25}) leads to the following equation for the frequency $W_k$: 
\begin{equation}\label{e28}
W_k^2=\Omega_k^2-\frac{1}{2}\left[\frac{W''_k}{W_k}-\frac{3}{2}
\frac{{W'_k}^2}{W_k^2}\right].
\end{equation}
This equation can be solved iteratively 
\begin{equation}\label{e29}
{W_k^{(N+1)}}^2=\omega_k^2-\Delta\xi a^2R-\frac{1}{2}
\left[\frac{{W_k^{(N)}}''}{{W_k^{(N)}}}-\frac{3}{2}
\frac{{{W_k^{(N)}}'}^2}{{W_k^{(N)}}^2}\right]
\end{equation}
with $W_k^{(0)}=\omega_k$ in the sense that for a finite time interval
and sufficiently large $k$ the RHS of eq. (\ref{e29}) is strictly
positive. Then, $W_k^{(N)}$ can be continued to all values of $k$ in
such a way that it is a smooth function of time.  As each iteration picks up 
two time derivatives the $N$th iterative solution $W_k^{(N)}$ is of
adiabatic order $2N$. Substituting $W_k^{(N)}$ back into (\ref{e27})
yields a so called  approximate adiabatic mode $\tilde{f}_k^{(N)}$.

An adiabatic vacuum state of iteration order $N$ is determined by a
complete set of mode solutions $\{f_k, f_k^*\}$ to eq. (\ref{e25})
satisfying initial conditions
\begin{equation}\label{e210}
f_k(\eta_0)=\tilde{f}_k^{(N)}(\eta_0), \quad 
f'_k(\eta_0)={\tilde{f}_k^{(N)^{\scriptstyle \prime}}}(\eta_0)\, ,
\end{equation}
i.e. an adiabatic mode coincides with an approximate mode 
$\tilde{f}_k^{(N)}$ on the initial Cauchy surface. With the particular form 
(\ref{e27}) of the approximate adiabatic modes these initial conditions 
read explicitly
\begin{equation}\label{e211}
f_k(\eta_0)=\frac{1}{\sqrt{2W_k^{(N)}(\eta_0)}}, \quad 
f'_k(\eta_0)=-\left(iW_k^{(N)}(\eta_0)+
\frac{{W_k^{(N)}}'(\eta_0)}{2W_k^{(N)}(\eta_0)}\right)f_k(\eta_0)\, .
\end{equation}
According to this construction an adiabatic vacuum state depends on
\begin{myitemize}
\item the initial time $\eta_0$
\item the order of iteration $N$,
\item the extrapolation of $W_k^{(N)}$ to small momenta $k$.
\end{myitemize}
In the following we simply write $W_k$ instead of $W_k^{(N)}$ for the 
adiabatic frequency. 

Varying the action with respect to the metric yields the energy-momentum tensor. 
For a real scalar field with arbitrary curvature coupling one finds \cite{bunch}
\begin{eqnarray}\label{e212}
T_{\mu\nu}&=&(1-2\xi)\partial_\mu\varphi\partial_\nu\varphi
             -2\xi \varphi \nabla_\mu\nabla_\nu\varphi 
       +(2\xi-\frac{1}{2})g_{\mu\nu}\partial^\rho\varphi\partial_\rho\varphi
       \nonumber\\
   && + 2\xi g_{\mu \nu} \varphi\Box \varphi -\xi G_{\mu \nu}\varphi^2 
      -\frac{1}{2}g_{\mu \nu}m^2 \varphi^2\, .
\end{eqnarray}
A mode sum representation of its (bare) expectation value is obtained by 
substituting the mode decomposition (\ref{e23}) into (\ref{e212}).
We choose the energy density and the trace as the two independent components.
They take the following form
\begin{eqnarray}\label{e213}
\langle T_{\phantom{0}0}^0\rangle\equiv \varepsilon&=& 
        \int \frac{d\mu(k)}{2\pi^2a^4}\left[
        3\Delta\xi\left(h'+2h^2\right)|f_k|^2-3\Delta\xi h\left(|f_k|^2\right)'
        \phantom{\frac{1}{2}}\right.\nonumber\\
        &&\left.\phantom{\int \frac{d\mu(k)}{2\pi^2a^2}}
        +\frac{1}{2}\left(|f'_k|^2+\Omega_k^2|f_k|^2\right)\right]\, ,\nonumber\\
\langle T_{\phantom{\mu}\mu}^\mu\rangle\equiv T&=&
        \int \frac{d\mu(k)}{2\pi^2a^4}\left[
        \left(6\Delta\xi h'+m^2a^2\right)|f_k|^2+6\Delta\xi h\left(|f_k|^2\right)'
        \right.\nonumber\\
        &&\phantom{ \frac{d}{2\pi^2a^4}}
        \left.\phantom{{|^2}'}
        -6\Delta\xi\left(|f'_k|^2-\Omega_k^2|f_k|^2\right)\right],
\end{eqnarray}
where the abbreviation $h=a'/a$ has been introduced. The measure $d\mu(k)$
implies integration over continuous and summation over discrete momenta
\begin{equation}\label{e214}
\int d\mu(k) = 
\begin{cases}
\int_0^\infty dk k^2 & \text{if}\quad \kappa=0,-1 \\
\sum_{k=1}^{\infty} k^2  & \text{if}\quad  \kappa=+1 \, .
\end{cases}
\end{equation}
We note that the dependence on the quantum state enters the expectation values
(\ref{e213}) via the initial conditions satified by the modes $f_k$. As we are
concerned with adiabatic states the modes $f_k$ satisfy the initial conditions
(\ref{e210}).

The formal expressions (\ref{e213}) are divergent and need to be 
renormalised. This task can be achieved by the method of adiabatic 
regularisation \cite{bunch,birrdav,andpark}. In this scheme the renormalised
energy momentum tensor is obtained by subtracting from the mode integrals
(\ref{e213}) their fourth order adiabatic expansion:
\begin{equation}\label{e215}
\langle T_{\mu\nu}\rangle_{ren}\stackrel{\rm def}{=}\langle T_{\mu\nu}\rangle-
\langle T_{\mu\nu}\rangle^{(4)}\, .
\end{equation}
This subtraction is to be interpreted as a renormalisation of the gravitational 
constant, the cosmological constant and the coupling constant of the squared
curvature term in the classical gravitational action. As it was shown
in
\cite{andpark} even for closed spatial geometry ($\kappa=+1$) the subtraction has 
to be performed with the continuum measure
\begin{equation}\label{e216}
\langle T_{\mu\nu}\rangle^{(4)}=\int_0^\infty dk\, \frac{k^2}{2\pi^2a^2}\, 
{\cal T}^{(4)}_{\mu\nu}
\end{equation}
in order to correctly reproduce the trace anomaly. The explicit form of the
subtraction terms ${\cal T}^{(4)}_{\mu\nu}$ can be found, e.g., in 
\cite{bunch,andpark}. Also, adiabatic regularisation has been shown to be 
equivalent to covariant point splitting \cite{andpark,birrell} and thus
results in an energy-momentum tensor satisfying Wald's axioms
\cite{fulling}.

\section{Initial states and the energy-momentum \\ tensor}\label{sec3}
\setcounter{equation}{0}

In this section we show that an adiabatic vacuum must be at least of 
order four for the expectation value of the energy-momentum tensor to be
non-singular on the initial Cauchy surface. Before proceeding with the
proof we wish to give an intuitive argument in order to illuminate the 
problem. 

Obviously, the subtraction procedure (\ref{e215}) only makes sense if
the ultraviolet divergences of the bare expressions are cancelled by the 
divergent terms of the adiabatic expansion, i.e. by all terms of 
${\cal T}^{(4)}_{\mu\nu}(\eta)$ up to $\omega_k^{-3}$. As the subtraction terms 
are local this cancellation has to occur at each instant of time. In other words, 
the bare expressions need to possess an asymptotic expansion for large momenta 
that reproduces the divergent terms of the adiabatic expansion {\it uniformly} with 
respect to time.  This includes in particular the initial time where the bare
expressions are directly given in terms of the initial conditions. The
simple idea is now to compare the asymptotic expansion of the
bare expressions for large $\omega_k$ with the divergent
part of the adiabatic expansion {\it at the initial time}.

With the adiabatic initial conditions (\ref{e211}) the expectation value of 
the energy-momentum tensor (\ref{e213}) at the initial time $\eta_0$ becomes:
\begin{eqnarray}\label{e31}
\varepsilon(\eta_0)&=& \int \frac{d\mu(k)}{2\pi^2a_0^4}
        \frac{1}{4}\left[W_{k0}+\frac{\Omega_{k0}^2}{W_{k0}}
        +\frac{\left({W_{k0}}'\right)^2}{4{W_{k0}}^3}
        +6\Delta\xi h_0\frac{{W_{k0}}'}{{W_{k0}}^2}
        \phantom{\frac{1}{2}}\right.\nonumber\\
        &&\left.\phantom{\int \frac{d\mu(k)}{2\pi^2a^2}}
        +6\Delta\xi\left(h_0'+2h_0^2\right)\frac{1}{W_{k0}}
        \right]\, ,\nonumber\\
T(\eta_0)&=&\int \frac{d\mu(k)}{2\pi^2a_0^4}
        \frac{1}{2}\left[
        \left(6\Delta\xi h'_0+m^2a_0^2\right)\frac{1}{W_{k0}}-6\Delta\xi h_0
        \frac{{W_{k0}}'}{{W_{k0}}^2}
        \right.\nonumber\\
        &&\phantom{ \frac{d}{2\pi^2a^4}}
        \left.\phantom{{|^2}'}
        -6\Delta\xi\left(W_{k0}-\frac{\Omega_{k0}^2}{W_{k0}}
        +\frac{\left({W_{k0}}'\right)^2}{4{W_{k0}}^3}\right)\right]\, ,
\end{eqnarray}
where the subscript $0$ indicates that the time argument of the respective quantity 
is set equal to the initial time $\eta_0$, i.e. $a_0\equiv a(\eta_0)$ etc.. The 
asymptotic expansion of the adiabatic frequency $W_k$ for large $\omega_k$ 
can be inferred from eq. (\ref{e29}) by induction in $N$
\begin{eqnarray}\label{e32}
W_k&=&\omega_k\left[1-\frac{q}{2\omega_k^2}(1-\delta_{N,0})
      -\frac{{M^2}''+q''+q^2}{8\omega_k^4}(1-\delta_{N,0})\right.\nonumber\\
&&\left.\phantom{\omega_kd}
+\frac{q''}{8\omega_k^4}(1-\delta_{N,0}-\delta_{N,1})+O(\omega^{-6})\right].
\end{eqnarray}
Then, the divergent terms of (\ref{e31}) are readily found
\begin{eqnarray}\label{e33}
\varepsilon(\eta_0)&=& \int \frac{d\mu(k)}{2\pi^2a_0^4}
        \left\{\frac{\omega_{k0}}{2}-\frac{q_0}{4\omega_{k0}}
        -\frac{q_0^2}{16\omega_{k0}^3}(1-\delta_{N,0})
        +3\Delta\xi h_0\frac{{M_0^2}'+q_0'\delta_{N,0}}{4\omega_{k0}^3}
        \phantom{\frac{1}{2}}\right.\nonumber\\
        &&\left.\phantom{\int \frac{d\mu(k)}{2\pi^2a^2}}
        +3\Delta\xi\left(h_0'+2h_0^2\right)\left[\frac{1}{2\omega_{k0}}+
        \frac{q_0}{4\omega_{k0}^3}(1-\delta_{N,0})\right]
        +O(\omega_{k0}^{-5})\right\},\nonumber\\
T(\eta_0)&=&\int \frac{d\mu(k)}{2\pi^2a_0^4}
        \left\{\left(6\Delta\xi h'_0+m^2a_0^2\right)\left[\frac{1}{2\omega_{k0}}+
        \frac{q_0}{4\omega_{k0}^3}(1-\delta_{N,0})\right]
        \right.\nonumber\\
        &&\phantom{ \frac{d}{2\pi^2a^4}}\left.\phantom{{|^2}'}
        -3\Delta\xi\frac{q_0}{\omega_{k0}}\delta_{N,0}
        -3\Delta\xi h_0\frac{{M^2_0}'+q_0'\delta_{N,0}}{2\omega_{k0}^3}
        \right.\nonumber\\
        &&\phantom{ \frac{d}{2\pi^2a^4}}\left.\phantom{{|^2}'}     
        +3\Delta\xi\frac{{M^2_0}''}{4\omega_{k0}^3}\left(1-\delta_{N,0}\right)
        +3\Delta\xi\frac{q_0''}{4\omega_{k0}^3}\delta_{N,1}
        +O(\omega_{k0}^{-5})\right\} .
\end{eqnarray}
We observe that the structure of the divergences in the energy density
coincides with that of the adiabatic expansion if $N>0$. For the trace, 
however this is only true if $N>1$ because the term proportional to $q_0''$ 
(being of adiabatic order four) only appears in the second and subsequent 
iterations in (\ref{e29}). So when subtracting the adiabatic expansion 
\cite{bunch,andpark} in the cases $N=0,1$ one is effectively introducing 
divergent terms that are not present at the initial moment and the momentum 
integrals do not exist (at the initial time $\eta_0$).

Even though this simple comparison shows the root of the problem it
only proves the necessity of the condition $N>1$ under the assumption
that the adiabatic expansion yields all the divergences present in
the theory and therefore has to be subtracted.  In order to give a
self-contained proof we have to show that $N>1$ is necessary for the
bare expressions to possess uniform (with respect to a finite time
interval, containing the initial time) large momentum asymptotic
behaviour that reproduces the divergent structure of the adiabatic
expansion. For this purpose we represent the adiabatic modes $f_k$ in
terms of a different set of mode solutions $g_k$, subject to the
conformal-like initial conditions
\begin{equation}\label{e34} 
g_k(\eta_0)=\frac{1}{\sqrt{2\Omega_k(\eta_0)}},\quad
g'_k(\eta_0)=-i\Omega_k(\eta_0) g_k(\eta_0)\, .
\end{equation}
As both mode solutions correspond to a homogeneous state they are related by a 
diagonal Bogoliubov transformation
\begin{equation}\label{e35}
f_k(\eta)=e^{i\phi_k}\left[\cosh\theta_k g_k(\eta)
          +e^{i\delta_k}\sinh\theta_kg_k^*(\eta)\right]\, .
\end{equation}
The identity $\cosh^2\theta_k-\sinh^2\theta_k=1$ ensures that the
normalisation constraint $f_k{f^*}'_k-f^*_kf'_k=i$ is preserved. The Bogoliubov
coefficients are determined by the initial conditions satisfied by the modes 
$f_k$ and $g_k$. Their particular combinations appearing in the
representation of the energy-momentum tensor are:
\begin{eqnarray}\label{e36}
\cosh 2\theta_k&=&\frac{1}{2}\left[\frac{\Omega_{k0}}{W_{k0}}+
\frac{W_{k0}}{\Omega_{k0}}+\frac{W_{k0}}{\Omega_{k0}}
\left(\frac{W'_{k0}}{2W_{k0}^2}\right)^2\right]\nonumber\\
\sinh 2\theta_k\cos\delta_k&=&\frac{1}{2}\left[\frac{\Omega_{k0}}{W_{k0}}-
\frac{W_{k0}}{\Omega_{k0}}-\frac{W_{k0}}{\Omega_{k0}}
\left(\frac{W'_{k0}}{2W_{k0}^2}\right)^2\right]\nonumber\\
\sinh 2\theta_k\sin\delta_k&=&-\frac{W'_{k0}}{2W_{k0}^2}\, .
\end{eqnarray}
The energy-momentum tensor (\ref{e213}) 
can now be expressed in terms of the modes $g_k$ and the Bogoliubov coefficients. 
As the problem of the initial singularity is less severe in the energy density we 
will show the following calculation only for the trace:
\begin{eqnarray}\label{e37}
T&=& \int \frac{d\mu(k)}{2\pi^2a^4}
        \left\{\left(6\Delta\xi h'+m^2a^2\right)
        \left[\cosh 2\theta_k|g_k|^2+\sinh 2\theta_k\Re
        \left(e^{-i\delta_k}g_k^2\right)\right]\right.\nonumber\\
  &&\left.-6\Delta\xi\left[\cosh 2\theta_k\left(|g'_k|^2-\Omega_k^2|g_k|^2\right)
    +\sinh 2\theta_k\Re\left(e^{-i\delta_k}\left({g'_k}^2-\Omega_k^2g_k^2
  \right)\right)\right]\right.\nonumber\\
  &&\left. +6\Delta\xi h\left[\cosh 2\theta_k(|g_k|^2)'+\sinh 2\theta_k\Re
    \left(e^{-i\delta_k}(g_k^2)'\right)\right]\right\}\, .
\end{eqnarray}
The next step consists in finding the large momentum behaviour of (\ref{e37}).
For this purpose we make use of an asymptotic expansion of the mode solutions $g_k$ 
that has similarily been used in \cite{baacke1,davunr,ringw}. The mode 
functions $g_k$ satisfy the oscillatory equation (\ref{e25}). Adding $\Omega_{k0}^2$ on 
both sides yields
\begin{equation}\label{e38}
g''_k+\Omega_{k0}^2g_k=-\left(\Omega_k^2-\Omega_{k0}^2\right)g_k
\equiv -\Delta\Omega^2g_k\, .
\end{equation}
The key point is that $\Delta\Omega^2$ is independent of $k$. Moreover, it
vanishes at the initial time: $\Delta\Omega^2(\eta_0)=0$. 
The quantity $\Omega^2_{k0}$ is strictly positive for sufficiently large momentum $k$
so that eq. (\ref{e38}) possesses the homogeneous solution 
$e^{-i\Omega_{k0}(\eta-\eta_0)}$. Then, with the help of the ansatz 
\begin{equation}\label{e39}
g_k(\eta)=\frac{e^{-i\Omega_{k0}(\eta-\eta_0)}}{\sqrt{2\Omega_{k0}}}
\left[1+\tilde{g}_k(\eta)\right]
\end{equation}
and using the initial conditions (\ref{e34}) the mode equation (\ref{e38})
can be transformed into the following integral equation 
\begin{equation}\label{e310}
\tilde{g}_k(\eta)=\frac{i}{2\Omega_{k0}}\int_{\eta_0}^{\eta}d\eta'
\left(e^{2i\Omega_{k0}(\eta-\eta')}-1\right)\Delta\Omega^2(\eta')
\left[1+\tilde{g}_k(\eta')\right]\, .
\end{equation}
This equation can be solved by iteration starting with $\tilde{g}_k^{(0)}\equiv 0$. 
As each iteration increases the power of $\Omega_{k0}^{-1}$ by one 
the iterative solution yields an expansion of $\tilde{g}_k$ in inverse powers of 
$\Omega_{k0}$ on the finite time interval $[\eta_0, \eta]$. The details of this 
expansion as well as the result for $\tilde{g}_k$ are displayed in the appendix.

It remains to derive the asymptotic expansion of the Bogoliubov parameters
(\ref{e36}) for large $\Omega_{k0}$. For this purpose we solve 
eq. (\ref{e28}) iteratively starting with $\Omega_k$ instead of $\omega_k$.
By induction in $\tilde{N}$ ($\tilde{N}$ is the number of iterations with respect to 
$\Omega_k$) we find:
\begin{eqnarray}\label{e311}
W_k^{(\tilde{N})}&=&
 \Omega_k\left[1-(1-\delta_{\tilde{N},0})\frac{{M^2}''}{8\Omega_k^4}
 +O(\Omega_k^{-6})\right],\nonumber\\
&=&\omega_k\left[1-\frac{q}{2\omega_k^2}
   -\frac{q^2+{M^2}''(1-\delta_{\tilde{N},0})}{8\omega_k^4}+O(\omega_k^{-6})\right],
\end{eqnarray}
where the second line is obtained by  means of $\Omega_k^2=\omega_k^2-q$.
The frequency $W_k^{(\tilde{N})}$ yields all terms up to $\omega_k^{-3}$ of a 
fourth order adiabatic frequency only if $\tilde{N}>0$ as can be seen by comparing 
(\ref{e311}) with eq. (\ref{e32}). 

With the help of relation (\ref{e311}) it is now straightforward to calculate the 
asymptotics of the Bogoliubov parameters (\ref{e36})
\begin{eqnarray}\label{e312}
\cosh 2\theta_k&=&1+O(\Omega_{k0}^{-6})\, ,\nonumber\\
\sinh 2\theta_k\cos\delta_k&=&(1-\delta_{\tilde{N},0})\frac{{M_0^2}''}{8\Omega_{k0}^4}
                              +O(\Omega_{k0}^{-6})\, ,\nonumber\\
\sinh 2\theta_k\sin\delta_k&=&-\frac{{M^2_0}'}{4\Omega_{k0}^3}
                              +O(\Omega_{k0}^{-5})\, .
\end{eqnarray}
Equipped with these expansions we finally isolate the divergent terms in the trace
of the energy-momentum tensor
\begin{eqnarray}\label{e313}
T(\eta)&=& \int \frac{d\mu(k)}{2\pi^2a^4}
        \left\{\left(6\Delta\xi h'+m^2a^2\right)
        \left(\frac{1}{2\Omega_{k0}}-\frac{\Delta\Omega^2}{4\Omega_{k0}^3}\right)
        -3\Delta\xi h\frac{{M^2}'}{2\Omega_{k0}^3}\right.\nonumber\\
  &&\left.+\frac{3\Delta\xi}{4\Omega_{k0}^3}\left[{M^2}''-\delta_{\tilde{N},0}{M_0^2}''
          \cos 2\Omega_{k0}(\eta-\eta_0)\right]+O(\Omega_{k0}^{-4})\right\}\, .
\end{eqnarray}
The term proportional to ${M_0^2}''\cos 2\Omega_{k0}(\eta-\eta_0)$ does not vanish
for $\tilde{N}=0$. Since the integral 
$\int d\mu (k) \Omega_{k0}^{-3}\cos 2\Omega_{k0}(\eta-\eta_0)$ 
diverges logarithmically in the limit $\eta\to \eta_0$ it leads to an initial 
singularity. All other divergent terms are indeed local and coincide with the 
divergence structure of the adiabatic expansion because we have
\begin{equation}\label{e314}
\frac{1}{\Omega_{k0}^3}=\frac{1}{k^3}+O(k^{-5})
\quad {\rm and} \quad
\frac{1}{\Omega_{k0}}-\frac{\Delta\Omega^2}{2\Omega_{k0}^3}=\frac{1}{k}
        -\frac{M^2}{2k^3}+O(k^{-5})
\, . 
\end{equation}
We conclude then, that the large momentum behaviour of the divergent terms of the
bare trace is uniform on the time interval $[\eta_0, \eta]$ only if $\tilde{N}>0$. 
In other words, an adiabatic vacuum state must be {\it at least of adiabatic order 
four} for the renormalised energy-momentum tensor to be finite on the initial Cauchy 
surface.
Therefore, only adiabatic states of order four or higher are reasonable physical
states. Some remarks are in order.

As the term causing the initial singularity is proportional to $\Delta\xi$ 
the problem of the dependence on the order of the adiabatic vacuum only affects
non-conformally coupled fields.

Since the expansion in inverse powers of $\Omega_{k0}$ reproduces the local
divergences of the adiabatic expansion one could ask why not subtract the leading terms 
of this expansion instead of the adiabatic ones? The answer
is that even though these subtractions are covariantly conserved they fail to 
reproduce the trace anomaly. To see this we rewrite the renormalised 
energy-momentum tensor (\ref{e215}) according to
\begin{equation}\label{e315}
\langle T_{\mu\nu}\rangle_{ren}=
\langle T_{\mu\nu}\rangle-\langle T_{\mu\nu}\rangle_{div}
+\langle T_{\mu\nu}\rangle_{div}-\langle T_{\mu\nu}\rangle^{(4)}
\end{equation}
and calculate the finite difference (with now $\tilde{N}>0$) 
\begin{equation}\label{e316}
\langle T_{\mu\nu}\rangle^{diff}
\equiv\langle T_{\mu\nu}\rangle_{div}-\langle T_{\mu\nu}\rangle^{(4)}\, ,
\end{equation}
where $\langle T_{\mu\nu}\rangle_{div}$ denotes all divergent terms of the 
inverse $\Omega_{k0}$ expansion (i.e. up to $\Omega_{k0}^{-3}$). 
The result can be represented as 
\begin{eqnarray}\label{e317}
T^{diff}&=&T^{Anomaly}-\frac{1}{8\pi^2}\left\{\left(m^4-m^2\Delta\xi R+
   3(\Delta\xi)^2\nabla^\mu\nabla_\mu R\right)\ln\frac{ma}{M_0}\right.\nonumber\\
 &&\left.+\frac{1}{a^2}\left(\frac{m^4}{4}g_{00}+m^2\Delta\xi G_{00}
   -\frac{1}{2}{}^{(1)}H_{00}\right)-\frac{3m^4}{4}
    +\frac{m^2}{36a^2}(1-18\Delta\xi)R\right.\nonumber\\
 &&\left.-\frac{1}{12}\Delta\xi\nabla^\mu\nabla_\mu R 
    -\frac{1}{4}(\Delta\xi)^2R^2
    -\frac{\kappa}{6a^4}\left[6\Delta\xi h'+m^2a^2(1-36\Delta\xi)\right]
    \right.\nonumber\\
 &&\left. +\frac{3}{a^2}(\Delta\xi)^2\left[2(h'+h^2)R+hR'\right]
      +\frac{M_0^2}{2a^4}(6\Delta\xi h'+m^2a^2)\right\}\, .
\end{eqnarray}
Here $G_{\mu\nu}$ is the Einstein tensor, the definition of ${}^{(1)}H_{\mu\nu}$ 
can be found, e.g., in \cite{birrdav}. 
$T^{Anomaly}$ is the anomalous trace \cite{christ}
\begin{eqnarray}\label{e318}
T^{Anomaly}&=&\lim_{m \to 0}\langle T_\mu^\mu\rangle_{ren}
             -\langle (\lim_{m \to 0}T_\mu^\mu)\rangle_{ren}\\
&=&-\frac{1}{2880\pi^2}\left[R^{\mu\nu}R_{\mu\nu}-\frac{1}{3}R^2
  +\left(30\xi-6\right)\nabla^\mu\nabla_\mu R +90(\Delta\xi)^2R^2\right]\, .
\nonumber
\end{eqnarray}
The energy density $\varepsilon^{diff}$ is calculated likewise. The
covariant conservation of $\langle T_{\mu\nu}\rangle^{diff}$ has
explicitly been checked. Besides the trace anomaly, $T^{diff}$
contains the logarithmic terms which give rise to the so-called anomalous
scaling as well as the renormalisation scale dependence \cite{lindig}.

So we see, that even though $\langle T_{\mu\nu}\rangle_{div}$ is
covariantly conserved and has the correct local singularity structure,
its subtraction does not yield the correct renormalised
energy-momentum tensor as it cannot reproduce the trace anomaly.

\section{Conclusions}

Since all adiabatic vacua are Hadamard states \cite{junker}, they are usually
considered physically admissible quantum states of linear Klein-Gordon fields 
on Robertson-Walker spaces. However, we find that the corresponding 
energy-momentum tensor developes a generic singularity on the initial Cauchy 
surface if the order of the adiabatic state is less than four. The divergent 
terms of the large momentum asymptotics of the energy-momentum tensor only 
coincide with those of the adiabatic expansion if the adiabatic vacuum is at 
least of order four.
As a result, an adiabatic vacuum state only results in an energy momentum tensor
satisfying Wald's axioms and thus is a {\it physically reasonable} state 
if it is {\it at least of order four}. 

This result is supported by the adiabatic particle picture developed in 
\cite{carmen}. There, this restricted class of adiabatic vacua is shown to lead to
a natural physical interpretation of the structure of the energy-momentum tensor. 
It splits into a local part (vacuum polarisation) containing all the divergences 
which have to be subtracted and a non-local piece due to particle creation. 

We have further shown that the subtraction of the divergent terms of the non-local 
large momentum expansion (\ref{e313}) (cf. the appendix) as suggested in 
\cite{baacke2} does not result in the correct renormalised energy-momentum tensor 
of a scalar field on a Robertson-Walker space because it fails to reproduce the 
trace anomaly. 
Nevertheless, this expansion can be useful in practical calculations of the 
energy-momentum tensor as the difference $\langle T_{\mu\nu}\rangle^{diff}$ 
between the divergent terms (\ref{e313}) and the adiabatic subtractions has been 
calculated explicitly (\ref{e317}). 
Only the remaining part needs to be calculated numerically.

\section*{Appendix}
\setcounter{equation}{0}
\renewcommand{\theequation}{A.\arabic{equation}}

In this appendix we wish to derive an asymptotic expansion of the conformal-like 
mode functions $g_k$ in inverse powers of $\Omega_{k0}$, i.e. for large momentum.
The Volterra-type integral equation (\ref{e310}) (which holds for sufficiently 
large $k$) serves us as the starting point. The iteration 
procedure
\begin{equation}\label{ea1}
\tilde{g}_k^{(n+1)}(\eta)=\frac{i}{2\Omega_{k0}}\int_{\eta_0}^{\eta}d\eta'
K_k(\eta,\eta')\left[1+\tilde{g}_k^{(n)}(\eta')\right]
\end{equation}
with $\tilde{g}_k^{(0)}(\eta)\equiv 0$ converges uniformly on the time interval
$[\eta,\eta_0]$ (for fixed $k$). According to (\ref{e310}) the kernel $K_k(\eta,\eta')$
is given by
\begin{equation}\label{ea2}
K_k(\eta,\eta')\stackrel{\rm def}{=}\left[e^{2i\Omega_{k0}(\eta-\eta')}-1\right]
\Delta\Omega^2(\eta')\, .
\end{equation}
As a result of the iteration, the solution $\tilde{g}_k(\eta)$ has the series
representation
\begin{equation}\label{ea3}
\tilde{g}_k(\eta)=\sum_{n=1}^{\infty}\left(\frac{i}{2\Omega_{k0}}\right)^n
\int_{\eta_0}^{\eta}d\eta_1 K_k(\eta,\eta_1)\cdots
\int_{\eta_0}^{\eta_{n-1}}d\eta_n K_k(\eta_{n-1},\eta_n)\, .
\end{equation}
The estimate
\begin{equation}\label{ea4}
\left|\tilde{g}_k(\eta)\right|\leq {\rm exp}\left\{\frac{1}{\Omega_{k0}}
\int_{\eta_0}^{\eta}d\eta'\left|\Delta\Omega^2(\eta')\right|\right\}-1
\end{equation}
shows that $\tilde{g}_k(\eta)$ remains bounded and goes to zero as $k\to \infty$.
An asymptotic expansion of $\tilde{g}_k(\eta)$ in inverse powers of $\Omega_{k0}$
can now be achieved by expanding each addend of the series (\ref{ea3}). 
For this purpose we provide repeatedly integration by parts ($\Delta\Omega^2$,
i.e. $R(\eta)$ is assumed to be smooth) to the most inner integral of the 
$n$th addend and find
\begin{eqnarray}\label{ea5}
&&\int_{\eta_0}^{\eta_{n-1}}d\eta_n K_k(\eta_{n-1},\eta_n)\,\,=\,\,
-\int_{\eta_0}^{\eta_{n-1}}d\eta_n \Delta\Omega^2(\eta_n)\\
&&\phantom{dddddd}-\sum_{m=0}^{\infty}\left(\frac{-i}{2\Omega_{k0}}\right)^{m+1}
\left[{\Delta\Omega^2}^{(m)}(\eta_{n-1})-{\Delta\Omega^2}^{(m)}(\eta_0)
{\rm e}^{2i\Omega_{k0}(\eta_{n-1}-\eta_0)}\right]\nonumber\, .
\end{eqnarray}
As all subsequent integrations have the same structure, they are treated likewise. 
The result is an asymptotic series for the $n$th addend of (\ref{ea3}) with 
leading term  
$$
\frac{1}{n!}
\left(\frac{-i}{2\Omega_{k0}}\int_{\eta_0}^{\eta}d\eta' \Delta\Omega^2(\eta')\right)^n
+O(\Omega_{k0}^{-(n+1)})\, .
$$
Consequently, all terms contributing to $\tilde{g}_k(\eta)$ up to order $\Omega_{k0}^{-n}$  
are contained in the first $n$ addends of (\ref{ea3}). 
If $n=4$ we find, for example,
\begin{eqnarray}\label{ea6}
\Re\tilde{g}_k(\eta)&=&
  -\frac{1}{4\Omega_{k0}^2}\left[\Delta\Omega^2+\frac{1}{2}I_1^2\right]
  +\frac{1}{8\Omega_{k0}^3}{\Delta\Omega^2_0}'\sin2\Omega_{k0}(\eta-\eta_0)
  \nonumber\\
&&+\frac{1}{16\Omega_{k0}^4}\left[{\Delta\Omega^2}''
  -{\Delta\Omega^2_0}''\cos2\Omega_{k0}(\eta-\eta_0)+{\Delta\Omega^2}'I_1
  +\frac{5}{2}\left(\Delta\Omega^2\right)^2\right.\nonumber\\
&&\left. +{\Delta\Omega^2_0}'I_1\cos2\Omega_{k0}(\eta-\eta_0)
  +\frac{1}{2}\Delta\Omega^2I_1^2+I_1I_2+\frac{1}{4!}I_1^4\right]
  +O(\Omega_{k0}^{-5})\nonumber\\
\Im\tilde{g}_k(\eta)&=&-\frac{1}{2\Omega_{k0}}I_1+\frac{1}{8\Omega_{k0}^3}
   \left[{\Delta\Omega^2}'-{\Delta\Omega^2_0}'\cos2\Omega_{k0}(\eta-\eta_0)
   \phantom{\frac{|}{|}}\right.\nonumber\\
&&\left.+\Delta\Omega^2I_1+I_2+\frac{1}{3!}I_1^3\right]-\frac{1}{16\Omega_{k0}^4}\left[
  {\Delta\Omega^2_0}''\sin2\Omega_{k0}(\eta-\eta_0)\right.\nonumber\\
&&\left.-{\Delta\Omega^2_0}'I_1\sin2\Omega_{k0}(\eta-\eta_0)\right]+O(\Omega_{k0}^{-5})
\, ,
\end{eqnarray}
where the abbreviation 
$$
I_m=\int^{\eta}_{\eta_0}d\eta'\left[\Delta\Omega^2(\eta')\right]^m
$$
has been used. Note that (\ref{ea6}) already contains all terms contributing 
to the divergences of the trace (\ref{e313}).

\section*{Acknowledgements}
I would like to thank C. Molina-Paris and S. Habib for stimulating discussions
and the astrophysics group of the Los Alamos National Laboratory for the 
warm hospitality extended to me.

\end{document}